\documentclass[aps,prb,twocolumn,showpacs,superscriptaddress,groupedaddress]{revtex4-1}
\usepackage{graphicx}
\usepackage{rotating}
\usepackage{bm}
\usepackage{amsmath} 
\usepackage{mathrsfs} 
\usepackage{amssymb}
\usepackage{color}

\newcommand{\eat}[1]{}

\begin{document} 
 \title{Hysteresis, reentrance, and glassy dynamics in systems of
   self-propelled rods} \author{Hui-Shun Kuan\textit{$^{a}$}, Robert
   Blackwell\textit{$^{b}$}, Loren E. Hough\textit{$^{b}$}, Matthew A. Glaser\textit{$^{b}$}, and M.
   D. Betterton\textit{$^{b}$}} \affiliation{\textit{$^{a}$Department
     of Chemistry and Biochemistry, University of Colorado at Boulder,
     Boulder, CO 80302, USA} \\\textit{$^{b}$Department of Physics,
     University of Colorado at Boulder, Boulder, CO 80302, USA} }
 \date{\today}

      \begin{abstract}
        Non-equilibrium active matter made up of self-driven particles
        with short-range repulsive interactions is a useful minimal
        system to study active matter as the system exhibits
        collective motion and nonequilibrium order-disorder
        transitions.  We studied high-aspect-ratio self-propelled rods
        over a wide range of packing fraction and driving to determine
        the nonequilibrium state diagram and dynamic properties.
        Flocking and nematic-laning states occupy much of the
        parameter space.  In the flocking state the average internal
        pressure is high and structural and mechanical relaxation
        times are long, suggesting that rods in flocks are in a
        translating glassy state despite overall flock motion.  In
        contrast, the nematic-laning state shows fluid-like behavior.
        The flocking state occupies regions of the state diagram at
        both low and high packing fraction separated by nematic-laning
        at low driving and a history-dependent region at higher
        driving; the nematic-laning state transitions to the flocking
        state for both compression and expansion. We propose that the
        laning-flocking transitions are a type of glass transition
        which, in contrast to other glass-forming systems, can show
        fluidization as density increases. The fluid internal dynamics
        and ballistic transport of the nematic-laning state may
        promote collective dynamics of rod-shaped microorganisms.
      \end{abstract}
      \pacs{87.10.Tf,64.60.Cn,05.65.+b}
\maketitle

Active matter made up of self-driven particles exhibits novel physical
properties include collective motion, nonequilibrium order-disorder
transitions, and anomalous fluctuations and mechanical response
\cite{ramaswamy10,*marchetti13}. Understanding active matter may aid
the development of new technologies including autonomously motile and
self-healing synthetic materials.  Examples of active matter include
animal flocks \cite{cavagna10}, crawling and swimming cells
\cite{rappel99,*cisneros11,zhang10a,thutupalli14}, vibrated granular
materials \cite{narayan07, deseigne10}, self-propelled colloidal
particles \cite{bricard13,palacci13}, and the cellular cytoskeleton
and cytoskeletal extracts \cite{nedelec97,*butt10}.

Among active matter, self-propelled rods (SPR) provide a useful
minimal model system. Self-propulsion and excluded volume interactions
via a short-range repulsive potential are the only ingredients; rod
alignment occurs through collisions. Experiments which may be
approximated as SPR include vibrated granular rods \cite{kudrolli08},
motion of cytoskeletal filaments on a motor-bound surface
\cite{butt10,schaller10}, and surface or film swarming of rod-like
bacteria \cite{sokolov07,zhang10a,wensink12a,thutupalli14}.  Because
of their simplicity SPR are attractive to simulation study
\cite{kraikivski06,peruani06,yang10,peruani11a,wensink12,wensink12a,mccandlish12,abkenar13}
and have also been the focus of analytic theory
\cite{baskaran08,*baskaran08a,wensink12}.
SPR display a rich variety of dynamic states, including collective
motion
\cite{vicsek95,*gregoire04,*chate08,*bertin06,*bertin09,*ginelli10,*peruani12,*aldana07,*ihle13,toner95,mishra10,gopinath12,peshkov12,weber13,deseigne10}
and formation of dynamic clusters
\cite{peruani06,schaller10,yang10,peruani11b,*peruani13,mccandlish12,abkenar13,weber13}.

For SPR, rod shape, density, and driving are important in determining
the dynamic behavior
\cite{peruani06,baskaran08,*baskaran08a,yang10,peruani11a,mccandlish12,wensink12,wensink12a,abkenar13}.
For low driving, equilibrium-like isotropic and nematic liquid crystal
phases are recovered \cite{baskaran08,*baskaran08a,abkenar13}.  For
higher driving, dynamic states characterized by the appearance of
flocks, stripes, and swirls appear
\cite{peruani06,baskaran08,*baskaran08a,mccandlish12,wensink12,wensink12a,abkenar13}.
Baskaran and Marchetti derived a hydrodynamic model from the kinetics
of SPR with two-rod collisions and determined a state diagram from
linear stability analysis of homogeneous states, finding that activity
lowers the isotropic-nematic transition density
\cite{baskaran08,*baskaran08a}.  Previous simulation work has observed
flocking and laning states similar to those we study
here\cite{wensink12,mccandlish12,abkenar13}, but did not measure on
dynamic state transitions, hysteresis, or structural and mechanical
properties. In this work, by studying the state diagram over a broader
range of parameters with extensive expansion and compression
simulations and mechanical and structural characterization, we
demonstrate strong hysteresis, the emergence of glassy dynamics in the
flocking state, and reentrant fluidization.

\begin{figure}[t]
\includegraphics[width=0.5 \textwidth]{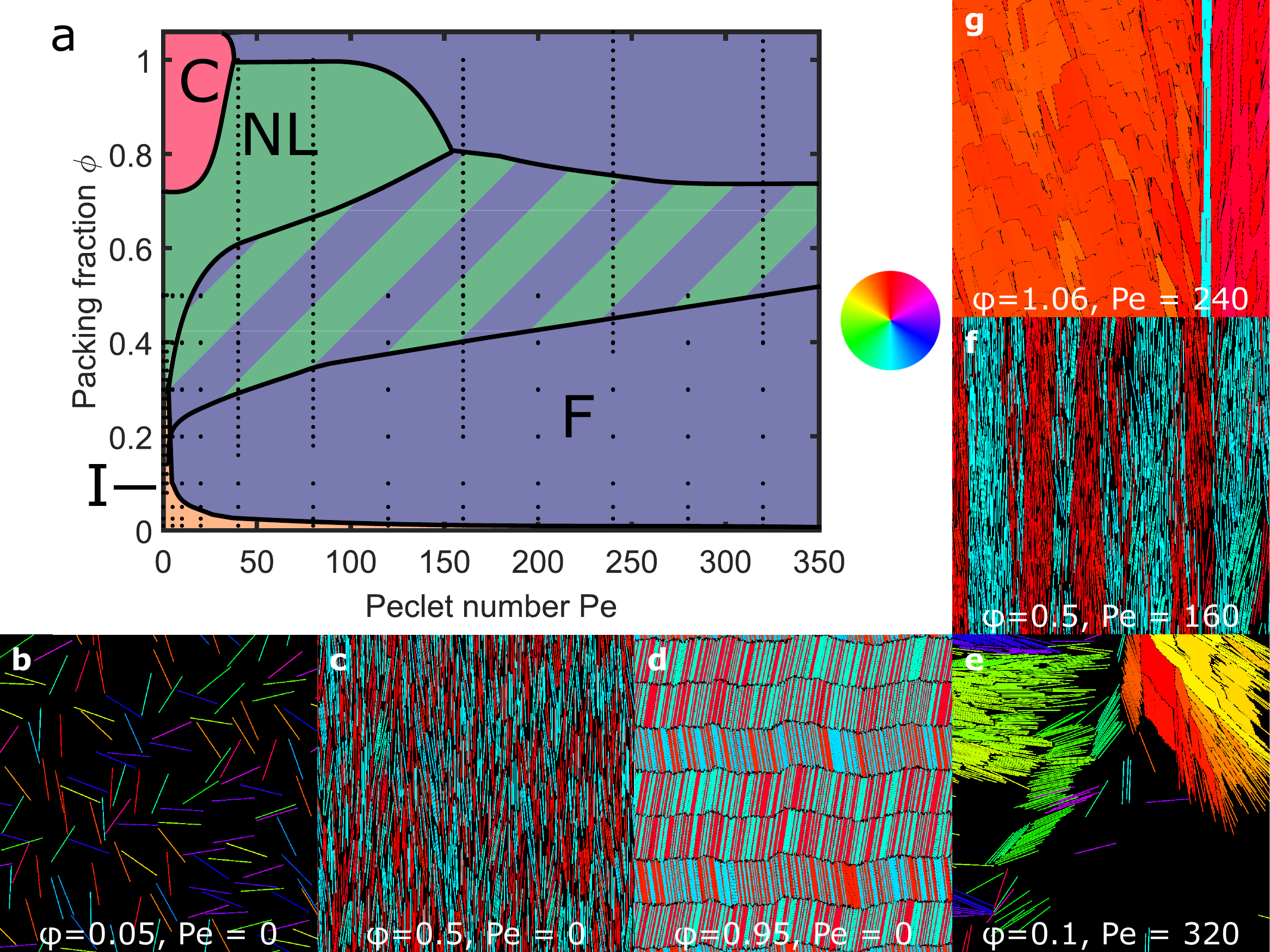}
\caption{Nonequilibrium state diagram and snapshots of self-propelled
  rods of aspect ratio 40.  (a) State diagram as a function of Peclet
  number and packing fraction. Points indicate parameter sets of
  simulations. Solid lines indicate boundaries of regions of stability
  of different initial conditions. I (orange), isotropic state; F
  (purple) flocking state; NL (green) nematic-laning state; C (pink)
  crystalline state. Green-purple striping indicates region where both
  flocking and nematic-laning initial conditions are stable. (b-g)
  Simulation snapshots at indicated packing fraction and Peclet
  number. Rods are colored by orientation according to the colorwheel.
  (b) Isotropic. (c) Nematic. (d) Crystal. (e) Flocking. (f)
  Nematic--laning. (g) Flocking.}
\label{overview}
\end{figure}

We studied self-propelled 2D spherocylinders with Brownian dynamics,
as in previous work \cite{mccandlish12}, using the computational
scheme of Tao et al.~\cite{tao05} developed for equilibrium
simulations of concentrated solutions of high-aspect-ratio particles.
Rods have length $L$ and diameter $\sigma$. The center-of-mass and
orientational equations of motion for rod $i$ with center-of-mass
position ${\bf r}_i$ and orientation ${\bf u}_i$ are
\begin{eqnarray}
  \label{eq:brownian}
  {\bf r}_i(t + \delta t) &=& {\bf r}_i(t) + {\bf \Gamma}_i^{-1}(t) \cdot
  {\bf F}_i(t) \delta t + \delta {\bf r}_i(t),\\
{\bf u}_i(t + \delta t) &=& {\bf u}_i(t) + {1 \over {\gamma_r}} {\bf
  T}_i(t) \times {\bf u}_i(t) \delta t + \delta {\bf u}_i(t),
\end{eqnarray}
where the random displacements $\delta {\bf r}_i(t)$ and
$\delta {\bf u}_i(t)$ are Gaussian-distributed,
${\bf \Gamma}_i^{-1}(t)$ is the inverse friction tensor, $\gamma_r$ is
the rotational drag coefficient, and ${\bf F}_i(t)$ and ${\bf T}_i(t)$
are the the deterministic force and torque on particle $i$
\cite{supplement}.  Excluded-volume interactions between particles are
modeled by the WCA potential as a function of the minimum distance
$s_{ij}$ between the two finite line segments of length $L$ that
define the axes of particles $i$ and $j$
\cite{supplement,weeks71}. The self-propulsion force is directed along
the particle axis with ${\bf F}_i^{\rm drive}= F_D {\bf u}_i$.  In the
absence of nonequibrium driving, this model has been
well-characterized both in 2D \cite{bates00} and 3D
\cite{bolhuis97,*mcgrother96}.

We nondimensionalize using the length $\sigma$, energy $k_B T$, and
time $\tau = D/\sigma^2$, where $D$ is the diffusion coefficient of a
sphere of diameter $\sigma$.  The three dimensionless parameters are
the rod aspect ratio $R = L/\sigma$, fixed at 40, the packing fraction
$\phi = A_{\rm rods}/A_{\rm system}$, and the translational Peclet
number ${\rm Pe} = F_D L/(k_B T)$.  We varied $\phi$ between $0.01$
and $1.04$ (where $\phi>1$ is possible due to the slight softness of
the repulsive potential), and Pe between 0 and 320.  We simulated
$N=4000$ rods in a square, periodic box. Most simulations were
initialized in an equilibrium isotropic, nematic, or crystalline
initial condition, then nonequilibrium activity was turned on and the
system was allowed to run for $10^7 \tau$. The simulation measurement
run was $10^7 \tau$, and the time step $\Delta t = 0.25 \tau$.

\begin{figure*}[t]
  \includegraphics[width=\textwidth]{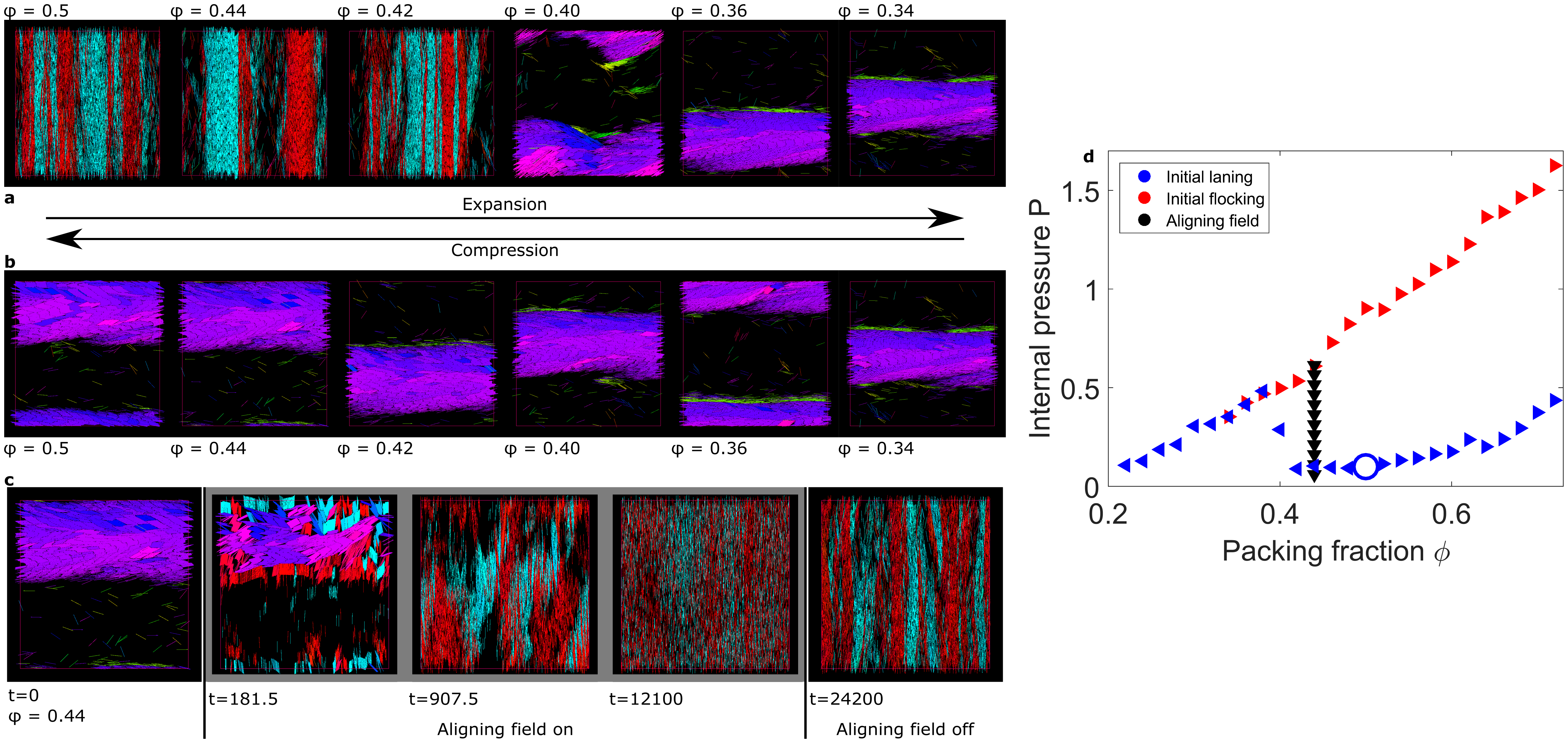}
  \caption{ Hysteresis in simulations under expansion, compression,
    and with an applied field. (a) Simulation snapshots of an
    expansion run initially in the nematic-laning state at $\phi=0.5$
    (far left). When the packing fraction reaches $0.4$ the system
    transitions to the flocking state. (b) Simulation snapshots of a
    compression run initially in the flocking state at $\phi=0.34$
    (far right). The system remains in the flocking state when
    compressed. (c) Simulation snapshots of run with applied nematic
    aligning field beginning in the flocking state at $\phi=0.44$ (far
    left). The applied field breaks up the flock and allows a return
    to the nematic-laning state after the field is removed (far
    right). (d) Internal pressure for systems shown in (a-c). Blue,
    initial nematic-laning state as shown in (a); red initial flocking
    state as shown in (b); black, system with nematic aligning field
    as shown in (c).}
\label{hysteresis}
\end{figure*}

At zero or low driving, we find equilibrium isotropic, nematic, and
crystalline states (fig.~\ref{overview}a-d). While we did not map the
equilibrium phase transitions in detail, our observations are
consistent with previous work \cite{bates00}.  As the Peclet number
increases, lower packing fractions roughly corresponding to the
equilibrium isotropic phase typically show flocking behavior
characterized by collective motion of clusters of various sizes
coexisting with a low-density vapor (fig.~\ref{overview}e), as
observed previously
\cite{peruani06,yang10,peruani11b,*peruani13,mccandlish12,weber13}.
While the flocking state remains globally isotropic (consistent with
previous predictions \cite{baskaran08}), the formation of dense
aligned clusters is characterized by short-range density correlations
that lead to peaks in the pair distribution function and the emergence
of polar and nematic orientational correlations that persist over a
cluster-size length scale (fig.~S1, and other data not shown).  Rod
mean-squared displacements are ballistic at short times, turning over
to diffusive at long times due to flock reorientation.  The long-time
angular mean-squared displacement is diffusive.

The flocking state shows large density heterogeneity suggestive of
two-phase coexistence between dense orientationally ordered clusters
and low-density isotropic rods. In previous work on self-propelled
spheres or disks, two-phase coexistence of a dense cluster and a
dilute vapor was observed that appears qualitatively similar to what
we observe here
\cite{palacci13,theurkauff12,*buttinoni13,henkes11,*fily12,*redner13,*speck14,*wensink14,*yang14,*takatori15,fily14}.
However, flocks are dynamic and are constantly merging, breaking up,
and exchanging particles with the dilute region
\cite{peruani06,peruani11b,*peruani13}. We identified flocks based on
measurements of the contact number
$c_i = \sum_{i \neq j} e^{-s_{ij}^2}$ and local polar order parameter
$p_i = \sum_{i \neq j}{\bf u}_i \cdot {\bf u}_j e^{-s_{ij}^2}/c_i$ of
rod $i$. Two-dimensional histograms show peaks in the density for
large $p_i$ over a range of $c_i$ (fig.~S2); individual flocks were
defined as collections of neighboring flock particles\cite{supplement}
(fig.~S2).  We identified flocks and isolated them in a box empty of
other rods; this led the isolated flock to break up, demonstrating
that flocks are not stable as isolated clusters.  Flock size
distributions are stable in time and power law in form with an
exponential cutoff, as observed
previously\cite{peruani06,zhang10a,peruani11,chen12,peruani13}
(fig.~S3).

As the Peclet number increases, higher packing fractions driven from
an equilibrium nematic or crystal typically show nematic-laning
behavior characterized by the formation of polar lanes of upward- and
downward-moving particles (fig.~\ref{overview}f,g). The density is
approximately uniform and the orientational order is globally nematic
in most cases with polar correlations on the scale of the system size
in the alignment direction and on the scale of a typical lane width
perpendicular to the alignment direction (fig.~S1 and data not
shown). Rod mean-squared displacements are ballistic in the alignment
direction and diffusive perpendicular, while the angular mean-squared
displacement is bounded due to the the maximum angular deviation of
rods.  The emergence of lanes in SPR and related models has been
observed in previous simulation studies
\cite{mccandlish12,wensink12,wensink12a,abkenar13,nagai15}, and laning
has been studied previously for spherical particles both in
experiments \cite{leunissen05,*sutterlin09,*vissers11} and
theory/simulation
\cite{chakrabarti03,dzubiella02,*netz03,*delhommelle05,*glanz12}.
Laning occurs because of the differences in collisions experienced by
rods as a function of their polar environment: a rod moving surrounded
by opposite polarity rods will experience more collisions, and
therefore more momentum transfer, than when surrounded by rods of
similar polarity. A rod surrounded by others of similar polarity will
therefore experience reduced lateral movement and be less likely to
leave the polar lane \cite{chakrabarti03,mccandlish12}.

To characterize the transitions between nematic-laning and flocking
states, we performed expansion and compression runs in which the
packing fraction was changed by $\Delta \phi = 0.02$, the simulation
was run for $10^7 \tau$ to reach a dynamic steady state, and then
measurements were performed over an additional $10^7 \tau$. The
appearance of the nematic-laning state is dependent on initial
conditions; lanes with equal numbers of up- and down-moving rods
result from initialization with an equilibrium nematic state and the
high rod packing fraction which prevents rod reorientation.  Upon
expansion, the system undergoes an abrupt transition to the flocking
state (fig.~\ref{hysteresis}a), while compression simulations
subsequently started in the flocking typically remain in the flocking
state (fig.~\ref{hysteresis}b). If we apply a nematic aligning field
to a compressed flocking state, the induced rod reorientation can
break up the flock and allow a transition back to the nematic-laning
state (fig.~\ref{hysteresis}c). This strong hysteresis is another
signature of an abrupt dynamic transition between the laning and
flocking states. While previous work has examined the nonequilibrium
state diagram of
SPR\cite{peruani06,baskaran08,*baskaran08a,mccandlish12,wensink12,wensink12a,abkenar13},
to our knowledge this is the first study to demonstrate strong
hysteresis in this system.

McCandlish et al.~found the laning state to be unstable to break up
\cite{mccandlish12}. While the strong hysteresis we observe makes it
difficult to guarantee that any nonequilibrum state is stable for
infinite time, our expansion and compression simulations effectively
extended our simulation times up to $2 \times 10^8 \tau$ in the
nematic-laning state, and upon reaching the transition boundary we
typically see break up of the lanes into flocks within the $10^7 \tau$
equilibration run. Therefore in our system the laning phase appears to
be stable, consistent with other work
\cite{wensink12,wensink12a,abkenar13}. The instablity observed by
McCandlish et al.~may be related to the reentrance we observe if the
simulations were performed near the upper limit of stability of the
nematic-laning state.

\begin{figure}[t]
\includegraphics[width=0.5 \textwidth]{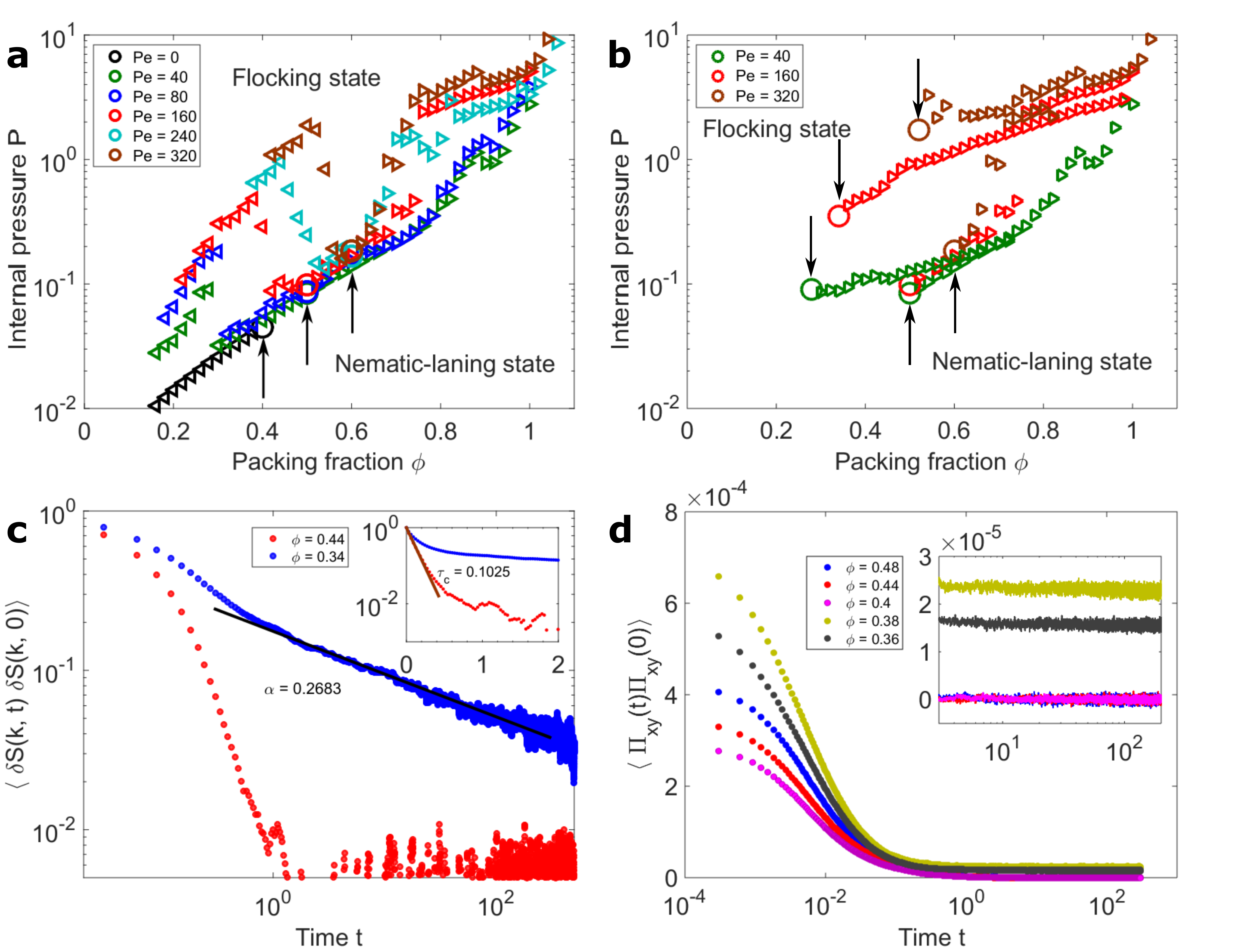}
\caption{Mechanical and structural properties of the nematic-laning
  and flocking states. (a,b) Internal pressure as a function of
  packing fraction during expansion (left-handed triangles) and
  compression (right-handed triangles) runs for simulations with
  varying Peclet number. The black arrows and open circles indicate
  the initial state of each run. (c) Structure-factor autocorrelation
  as a function of time for Pe=160 and systems in the laning
  ($\phi=0.44$, red) and flocking ($\phi = 0.34$, blue) states for the
  peak nearest wave number $k=2 \pi/\sigma$. The autocorrelation
  exhibits exponential decay in the nematic-laning state with
  characteristic time $\tau_c=0.10$ and power-law decay in the
  flocking state with the exponent $\alpha=0.27$ as indicated. Inset,
  semi-log plot. (d) Stress-tensor autocorrelation as a function of
  time for Pe=160 and systems in the nematic-laning ($\phi=0.4-0.48$)
  and flocking ($\phi = 0.36-0.38$) states. Inset, zoomed view of
  long-time tail.}
\label{pressure}
\end{figure}

During expansion runs, the isotropic internal pressure $P_o$, measured
by the virial\cite{supplement}, abruptly changes by a factor of 2--10
at the transitions between nematic-laning and flocking states
(fig.~\ref{hysteresis}d). (The nature of the pressure in active
systems has been the subject of recent work
\cite{solon14,*solon14a,*takatori14}; here we consider the internal
pressure determined by the virial only.)
At the highest packing fractions the internal pressure approaches a
plateau value near 10 for all systems, suggesting that a pure dense
flocking state has been reached. The internal pressure of the flocking
state lies along an envelope that decreases with decreasing packing
fraction as the rod flocking/isotropic fraction varies. Nematic-laning
systems undergo transitions to flocking upon both expansion and
compression (fig.~\ref{pressure}a,b, open circles labeled by arrows
indicate starting simulations of expansion/compression runs). Flocking
systems, typically remain flocking upon compression, but for low
packing fractions a transition back to the nematic-laning state upon
compression can occur (fig.~\ref{pressure}b, open circles labeled by
downward-pointing arrows indicate starting simulations of compression
runs).

The dense clusters and high pressure in the flocking state suggest
that the clusters may have slow internal dynamics.  To characterize
structural relaxation we measured the normalized structure-factor
autocorrelation function $C(t)/C(0)$, where
$C(t)=\langle \delta S(k,t) \delta S(k,0) \rangle$, $k$ is the
magnitude of the wavevector and
$\delta S(k,t) = S(k,t) - \left\langle S(k,t) \right\rangle$ is the
fluctuation in the the angle-averaged structure factor
$S(k,t) = {{1} \over {2 \pi N}} \int_0^{2 \pi} d\phi \rho({\bf k}, t)
\rho(-{\bf k}, t) $
\cite{supplement}.  Because the angle-averaged structure factor is
rotationally invariant, its autocorrelation probes internal structural
relaxation of flocks and lanes but is insensitive to flock
reorientation.  We determined the location of the peak nearest to wave
number $k=2 \pi/\sigma$, corresponding to side-by-side filaments
separated by approximately one diameter. In the nematic-laning state,
the structure-factor autocorrelation exponentially decays
(fig.~\ref{pressure}c, red curve). However in the flocking state, the
structure-factor autocorrelation has a power-law tail, indicating slow
structural relaxation (fig.~\ref{pressure}c, blue curve). Expansion to
lower packing fractions has little effect on the power-law exponent,
indicating that slow relaxation of dense clusters controls the decay
of the structure-factor autocorrelation.  Compression leads to a
density-dependent exponent (fig.~S4).

Mechanical relaxation was measured by the autocorrelation function of
the off-diagonal internal stress tensor
$\langle \Pi_{xy}(t) \Pi_{xy}(0) \rangle$ \cite{supplement}.  In the
nematic-laning state, the stress autocorrelation drops to zero around
$t = 1$ (fig.~\ref{pressure}d, blue, red, and purple curves). In the
flocking state, the stress autocorrelation function relaxes to a small
but long-lived plateau (fig.~\ref{pressure}d, yellow-green and grey
curves).  Consistent with this, the effective shear viscosity measured
via the Green-Kubo relation \cite{supplement} shows a factor of $10^3$
increase upon transition from the nematic-laning to the flocking state
for Pe=80 (fig.~S4).

The large increases in pressure and shear viscosity and slowed
structural and mechanical relaxation that occurs upon transition from
nematic-laning to flocking suggest that this is a type of glass
transition in which flocks, although collectively moving, have an
internally glassy, solid-like structure. Related observations were
made in an experimental system with self-propelled colloids, for which
nonequilibrium driving promoted formation of small, mobile crystalline
clusters \cite{palacci13}.  Related phase separation between a
low-density gas and high-density liquid, glassy clusters or crystals
has been observed both in experiments
\cite{palacci13,theurkauff12,*buttinoni13} and theory and simulations
\cite{henkes11,*fily12,*redner13,*speck14,*wensink14,*yang14,*takatori15,fily14}.
In contrast to both recent active jamming work and classic granular
jamming \cite{liu98,*reichhardt14}, in our self-propelled rod system
the increased importance of aligning interactions means that the
transition to the translating glassy flocking state can occurs both as
density is raised and \textit{lowered}. This reentrant fluidization
appears to be a novel feature of this transition in systems of
self-propelled rods. 

\begin{figure}[t]
\includegraphics[width=0.5 \textwidth]{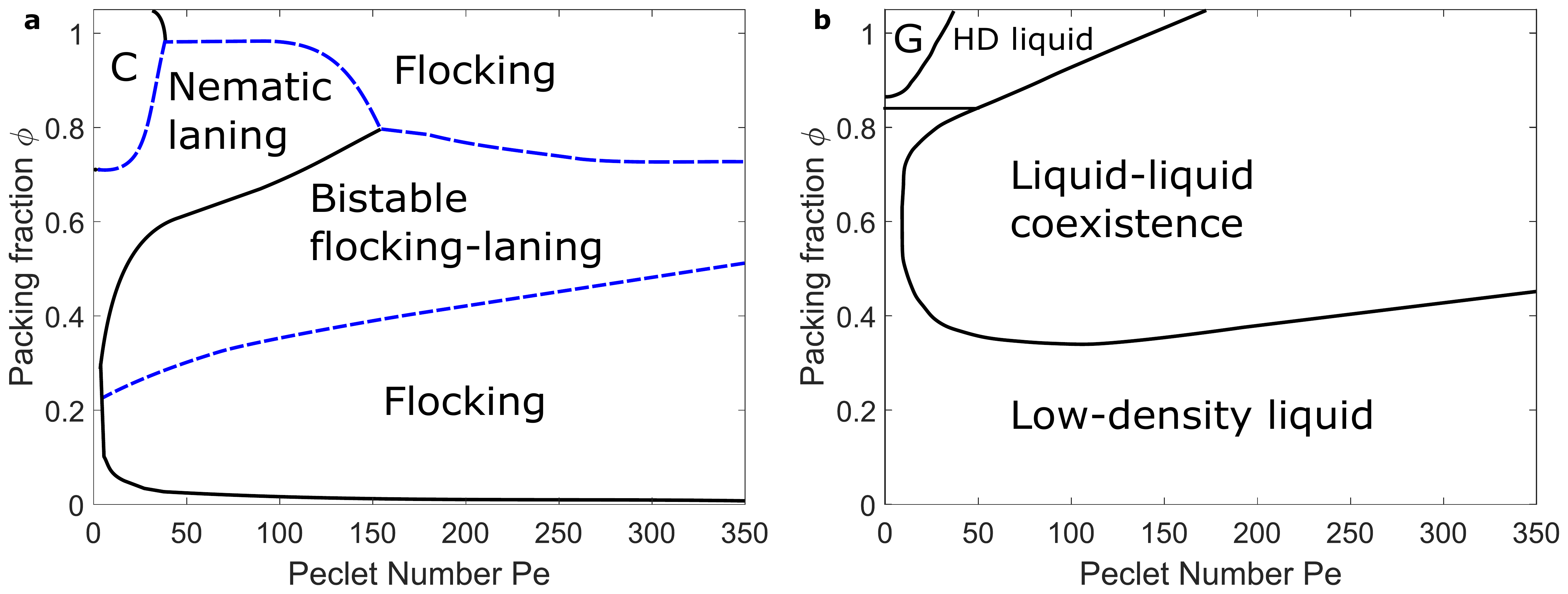}
\caption{Comparison of self-propelled rod and sphere nonequilibrium
  state diagrams. (a) SPR state diagram from this study as a function
  of the translational Peclet number. Blue dashed lines show limits of
  stability of the nematic-laning state characterized by ballistic
  transport along the lane. (b) Self-propelled sphere state diagram as
  a function of the rotational Peclet number, adapted from Fily,
  Henkes, and Marchetti \cite{fily14}. }
\label{compare}
\end{figure}

Self-propelled rods couple shape anisotropy to directional polarity,
in contrast to self-propelled spheres. This enables a rich state
diagram for SPR with important implications for transport
(fig.~\ref{compare}). Orientational ordering allows SPR to form a
nematic-laning state at high packing fraction characterized by fluid
internal dynamics and ballistic transport along the lanes.  Much of
the same region of parameter space of self-propelled spheres consists
of phase-separated liquid-liquid coexistence
(fig.~\ref{compare})\cite{henkes11,*fily12,*redner13,*speck14,*wensink14,*yang14,*takatori15,fily14}
for which particle dynamics are diffusive\cite{fily14} and the
formation of dense clusters limits particle motion. Perhaps the
physics of laning is important for collective motion of rod-shaped
microorganisms such as \textit{Myxococcus xanthus}, which during
fruiting-body formation assemble into dense streams qualitatively
similar to the lanes we observe\cite{thutupalli14}. Ballistic
transport through coupling of orientational order and self propulsion
may give an advantage to rod-shaped rather than spherical bacteria.

\begin{acknowledgments}
  We thank Lisa Manning, John Toner, Leo Radzihovsky, and Joel Eaves
  for useful discussions. This work was supported by NSF grants
  MRSEC-DMR-0820579, EF-ATB-1137822, and DMR-0847685 and NIH grant T32
  GM-065103.  This work utilized the Janus supercomputer, which is
  supported by the National Science Foundation (CNS-0821794) and the
  University of Colorado Boulder. The Janus supercomputer is a joint
  effort of the University of Colorado Boulder, the University of
  Colorado Denver and the National Center for Atmospheric Research.
  Janus is operated by the University of Colorado Boulder.
\end{acknowledgments}

\bibliography{collective,zoterolibrary}{}
\bibliographystyle{apsrev4-1}

\end{document}